\documentclass[aps,prb,nobibnotes,amsmath,amssymb,superscriptaddress,a4paper,twocolumn, showkeys]{revtex4-2}

\setcitestyle{super}
\usepackage[utf8]{inputenc}
\usepackage{amsmath}
\usepackage{amsfonts}
\usepackage{amssymb}
\usepackage{graphicx}
\usepackage{textgreek}
\usepackage{miller}
\usepackage{physics}
\usepackage[version=4]{mhchem}
\usepackage{xcolor} 

\makeatletter
\newcommand{\fmarki}{*}
\newcommand{\fmarkii}{\ensuremath{\dagger}}
\newcommand{\fmarkiii}{\ensuremath{\ddagger}}
\newcommand{\fmarkiv}{\ensuremath{\mathsection}}
\newcommand{\fmarkv}{\ensuremath{\mathparagraph}}
\newcommand{\fmarkvi}{\ensuremath{\|}}
\newcommand{\fmarkvii}{**}
\newcommand{\fmarkviii}{\ensuremath{\dagger\dagger}}
\newcommand{\fmarkix}{\ensuremath{\ddagger\ddagger}}
                
\def\@fnsymbol#1{{\ifcase#1\or \fmarki\or \fmarkii\or \fmarkiii\or \fmarkiv\or \fmarkv\or \fmarkvi\or \fmarkvii\or \fmarkviii\or \fmarkix \else\@ctrerr\fi}}
\makeatother

\renewcommand{\fmarki}{\ensuremath{\dagger}}
\renewcommand{\fmarkii}{*}

\begin{document}


\author{Dags Olsteins}
\altaffiliation{Authors with equal contribution}
\affiliation{Department of Energy Conversion and Storage, Technical University of Denmark, 2800 Kgs.Lyngby, Denmark}

\author{Gunjan Nagda}
\altaffiliation{Authors with equal contribution}
\affiliation{Center For Quantum Devices, Niels Bohr Institute, University of Copenhagen,\\ 2100 Copenhagen, Denmark}

\author{Damon J. Carrad}
\affiliation{Department of Energy Conversion and Storage, Technical University of Denmark, 2800 Kgs.Lyngby, Denmark}

\author{Daria V. Beznasyuk}
\affiliation{Department of Energy Conversion and Storage, Technical University of Denmark, 2800 Kgs.Lyngby, Denmark}

\author{Christian E. N. Petersen}
\affiliation{Department of Energy Conversion and Storage, Technical University of Denmark, 2800 Kgs.Lyngby, Denmark}

\author{Sara Martí-Sánchez}
\affiliation{Catalan Institute of Nanoscience and Nanotechnology (ICN2), CSIC and BIST, Campus UAB, 08193 Bellaterra, Barcelona, Catalonia, Spain}

\author{Jordi Arbiol}
\affiliation{Catalan Institute of Nanoscience and Nanotechnology (ICN2), CSIC and BIST, Campus UAB, 08193 Bellaterra, Barcelona, Catalonia, Spain}
\affiliation{ICREA, Passeig de Lluís Companys 23, 08010 Barcelona, Catalonia, Spain}

\author{Thomas Sand Jespersen}
\email[Corresponding author email: ]{tsaje@dtu.dk}
\affiliation{Department of Energy Conversion and Storage, Technical University of Denmark, 2800 Kgs.Lyngby, Denmark}
\affiliation{Center For Quantum Devices, Niels Bohr Institute, University of Copenhagen,\\ 2100 Copenhagen, Denmark}

\title{Statistical Reproducibility of Selective Area Grown InAs Nanowire Devices}



\begin{abstract}

\textbf{Abstract:} New approaches such as selective area growth (SAG), where crystal growth is lithographically controlled, allow the integration of bottom-up grown semiconductor nanomaterials in large-scale classical and quantum nanoelectronics. This calls for assessment and optimization of the reproducibility between individual components. We quantify the structural and electronic statistical reproducibility within large arrays of nominally identical selective area growth InAs nanowires. The distribution of structural parameters is acquired through comprehensive atomic force microscopy studies and transmission electron microscopy. These are compared to the statistical distributions of the cryogenic electrical properties of $256$ individual SAG nanowire field effect transistors addressed using cryogenic multiplexer circuits. Correlating measurements between successive thermal cycles allows distinguishing between the contributions of surface impurity scattering and fixed structural properties to device reproducibility. The results confirm the potential of SAG nanomaterials, and the methodologies for quantifying statistical metrics are essential for further optimization of reproducibility.

\end{abstract}

\maketitle


The characteristics of nanoscale electrical devices can be significantly influenced by rearrangements of only a few atoms in the vicinity of the active transport channel, especially at low temperatures. \cite{riel:2014,delalamo:2011} Material purity and device reproducibility are thus key concerns in the development of increasingly complex and powerful electrical quantum circuits operating at ultra-low temperatures.\cite{gonzalez-zalba:2021} While bottom-up semiconductor nanowires have been an important platform for mesoscopic quantum electronics due to the high control of crystal properties,\cite{dick:2010} intrinsic quantum confinement,\cite{ford:2009} and flexible epitaxial integration of dissimilar materials,\cite{bjork:2002,krogstrup:2015} the traditional out-of-plane geometry is incompatible with standard semiconductor processing and has prevented up-scaling of nanowire circuits.
The method of SAG\cite{Wang2019Jun,op2020plane, Raya2020Jan, Bollani2020Jan} potentially removes this roadblock by allowing large-scale lithographic control of planar nanowire growth. Already, key device concepts based on SAG nanowires have been realized, such as field effect transistors (FETs), \cite{beznasyuk:2022,Friedl2018Apr} nanowire Hall bars, \cite{Lee2019Aug,seidl2021postgrowth} quantum interferometers, \cite{vaitiekenas2018selective,Krizek2018Sep} hybrid superconducting devices,\cite{hertel:2022, vaitiekenas2018selective, goswami:2023} and quantum dots.\cite{ten2022large} So far, however, the focus has been on proof-of-principle experiments based on a few devices. We here take the next step towards up-scaling by quantifying the statistical reproducibility of SAG devices operated at the deep cryogenic conditions relevant for quantum circuitry.

Statistical assessment of reproducibility requires repeated characterization of the electrical parameters for large ensembles of devices. To bypass the device-count limitation usually imposed on cryogenic nanowire experiments, we employ cryogenic multiplexers (MUX) for semi-automated characterization of large numbers of devices in a single cool-down.\cite{smith:2014,smith2020high,olvsteins2023cryogenic} The variability between nominally identical devices is influenced by both structural variations related to fabrication tolerances and local fluctuations in growth conditions and from the random electrostatic potential from charged impurities or adsorbates in the vicinity of the device. We study these relative contributions by statistically correlating measurements of successive temperature cycles between the cryogenic regime and room temperature, where impurity configurations are randomized.\cite{see:2012} These distributions are compared to the fixed structural variability that is quantified through systematic atomic force microscopy (AFM) analysis of $180$ nominally identical nanowires.

\begin{figure*}[htb!]
  \centering
    \includegraphics[width = \linewidth,page=1]{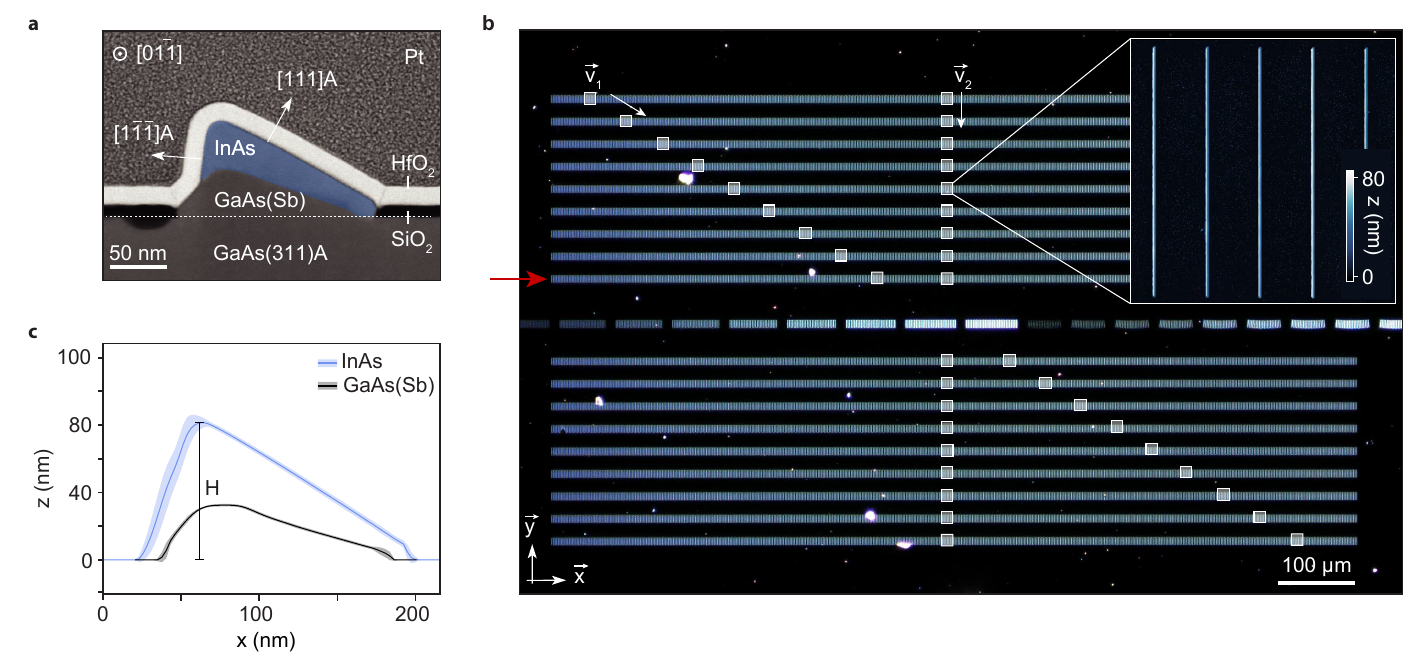} 
    \caption{\textbf{a} Cross-section transmission electron micrograph of a SAG InAs nanowire (blue). The asymmetric shape is a consequence of the \hkl(311) substrate symmetry. \textbf{b} Optical dark-field microscope image of a nanowire array consisting of two blocks of $9$ rows, each holding $512$ individual, nominally identical SAG nanowires with a length of $10\,\mu \mathrm m$ and cross-sections as in panel a. $180$ nanowires were structurally characterized by AFM acquired at the positions of the white squares. An example of an AFM image is shown in the inset. The red arrow indicates the row of nanowires used in the transport measurements. \textbf{c} Example of cross-section AFM profile averaged over $1 \mu \mathrm m$ of an InAs nanowire and the GaAs(Sb) buffer. The shaded regions correspond to the standard deviation. The maximum height, $H$, is used to quantify the morphological variations in the nanowire array.
    \label{Fig1}}
  \hfill
\end{figure*}

The aim of this work is thus to quantify the statistical variations in the specific SAG InAs nanowires grown here, which serves as a reference for ongoing efforts to optimize SAG as a scalable platform for quantum electronics. Secondly, our work demonstrates the methods for acquiring the required statistical metrics in SAG circuits. We compare the results to alternative material candidates for cryogenic electronics, such as AlGaAs/GaAs 2D heterostructures and cryogenic CMOS.


SAG nanowires were grown by molecular beam epitaxy (MBE) on undoped \hkl(311)A GaAs substrates, which are electrically insulating at low temperatures. A growth mask ($10$\,nm of $\mathrm{SiO}_2$) was patterned using electron beam lithography and dry etching to expose the GaAs substrate in large arrays of nominally identical $0.15 \times 10\,\mu \mathrm{m}$ rectangular openings aligned along the $[0\overline{1}1]$ direction. The arrays contain two blocks. Each block consists of $9$ rows spaced $35 \, \mu \mathrm m$ apart, and holds $512$ nanowires with a pitch of $2 \,\mu \mathrm{m}$. The prepared substrates were introduced into the MBE chamber, and a GaAs(Sb) buffer layer was grown to improve the GaAs substrate surface for the subsequent growth of InAs which serves as the active conducting channel in the transport experiments.\cite{beznasyuk:2022,olvsteins2023cryogenic} Further details of substrate preparation and growth parameters are included in Supplementary Section I. Figure \ref{Fig1}\textbf{a} shows a high-angle annular dark field scanning transmission electron microscope (HAADF STEM) micrograph of a SAG nanowire cross-section. The nanowire has an asymmetric profile imposed by the \hkl(311) symmetry of the substrate. Figure \ref{Fig1}\textbf{b} shows a dark-field optical microscope image of an array consisting of $8192$ nanowires. Three such arrays from the same growth were used in this study for the TEM, AFM, and transport characterization.

\begin{figure}[htb]
  \centering
    \includegraphics[width = 8.5 cm,page=1]{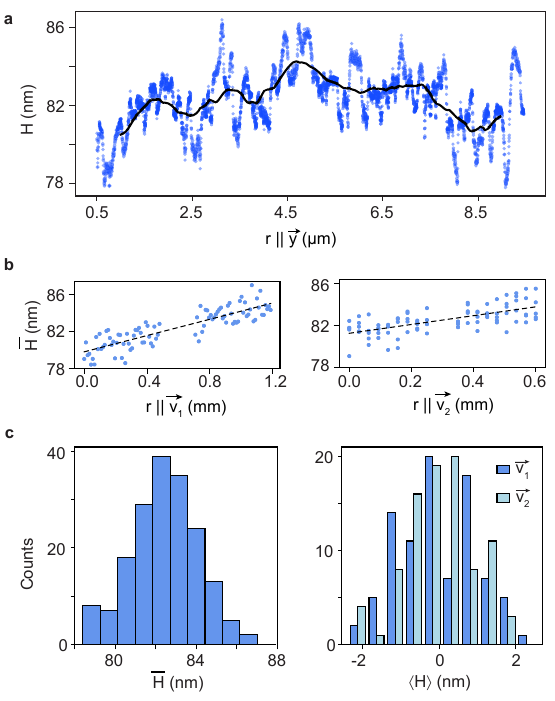} 
    \caption{\textbf{a} The height, $H$, defined in Fig.\ 1\textbf{c} along a typical SAG nanowire in the array of Fig.\ 1. The black curve shows the running average over $1\,\mu \mathrm m$. \textbf{b} $H$ averaged over a middle $1\,\mu \mathrm m$ segment for nanowires located along $\Vec{v}_1$ ($\Vec{v}_2$) as defined in Fig.\ 1\textbf{b}. The dashed lines are linear fits. \textbf{c} Histograms of the average $\overline H$ and the corresponding residuals $\langle H \rangle$ after subtracting linear fits to data in panels b, quantifying short-range variation between wires of the array. 
    } 
    \label{Fig2}
  \hfill
\end{figure}

We first consider the contributions to structural variations between neighboring nanowires and across the array. Although all nanowires are defined to be nominally identical, different effects contribute to variations at different length scales. While variations that occur at length scales significantly below the typical device length ($L_\mathrm{D} \sim 1 \, \mu \mathrm m$, see below) alter the average electrical parameters, the variations at longer scales reduce reproducibility between neighboring devices or lead to long-range modulations across the array. For example, on the few-nanometre scale, dislocations and stacking faults occur in the crystal due to the lattice mismatch at the GaAs/InAs interface. This is confirmed by cross-sectional TEM performed on four representative nanowires (see Supplementary Section III). Elastic relaxation is observed around the rounded top corner and the shorter \hkl(1-1-1)A nanowire facet, while misfit dislocations occur with an average distance of $\sim 9 \, \mathrm{nm}$ along the longer \hkl(111)A InAs/GaAs(Sb) interface. In addition, based on a limited number of HR-TEM micrographs, a typical number of stacking faults between 1 and 4 are observed in each cross-section. As discussed previously,\cite{beznasyuk:2022} the strain in the system and the elevated temperature during growth promote inhomogeneous diffusion of Ga from the GaAs(Sb) buffer layer during InAs growth as confirmed in cross-sectional HAADF STEM and electron energy loss spectroscopy (EELS) (Supplementary Sections III and IV). Such material intermixing will affect the electrical properties locally due to associated distortions of the band structure or defect states. As dislocations and stacking faults occur on a scale much shorter than $L_\mathrm{D}$ we expect that they mainly affect average electrical parameters such as the mobility, and contribute less to device-to-device variation. The impact of stacking faults on the mobility has been shown previously for out-of-plane InAs nanowires\cite{schroer:2010}, and a significantly improved mobility was observed in Ref.\ \cite{beznasyuk:2022} by suppressing the formation of crystal faults.

On an intermediate scale ($\sim 20-100$\,nm), etch roughness of the lithographically defined growth apertures and effects of random growth nucleation on the SiO$_2$ mask can lead to local structural variations along nanowires. Even larger scale modulations across the array ($\mu \mathrm{m - mm}$) could arise from spatial variations in substrate temperature and adatom flux density across the substrate. To quantify these effects, detailed topographic AFM maps were acquired of $180$ individual nanowires located at the positions of the white squares in Fig.~\ref{Fig1}\textbf{b}. An example of the average profiles of a $1\,\mu \mathrm m$ long segment from a complete nanowire structure, and from a GaAs(Sb) reference growth are shown in Fig.~\ref{Fig1}\textbf{c}, where the colored bands represent the standard deviation. As a characteristic parameter quantifying the morphological variations along nanowires and between nanowires of the array, an automatic procedure was developed to extract the maximum height, $H$, at each measured point along each nanowire (See Supplementary Section V). As the cross-sectional shape is defined by the \hkl[111] and \hkl[1-1-1] crystal planes, other dimensions scales with $H$..

\begin{figure*}[htb]
    \centering
    \includegraphics[width = 16 cm]{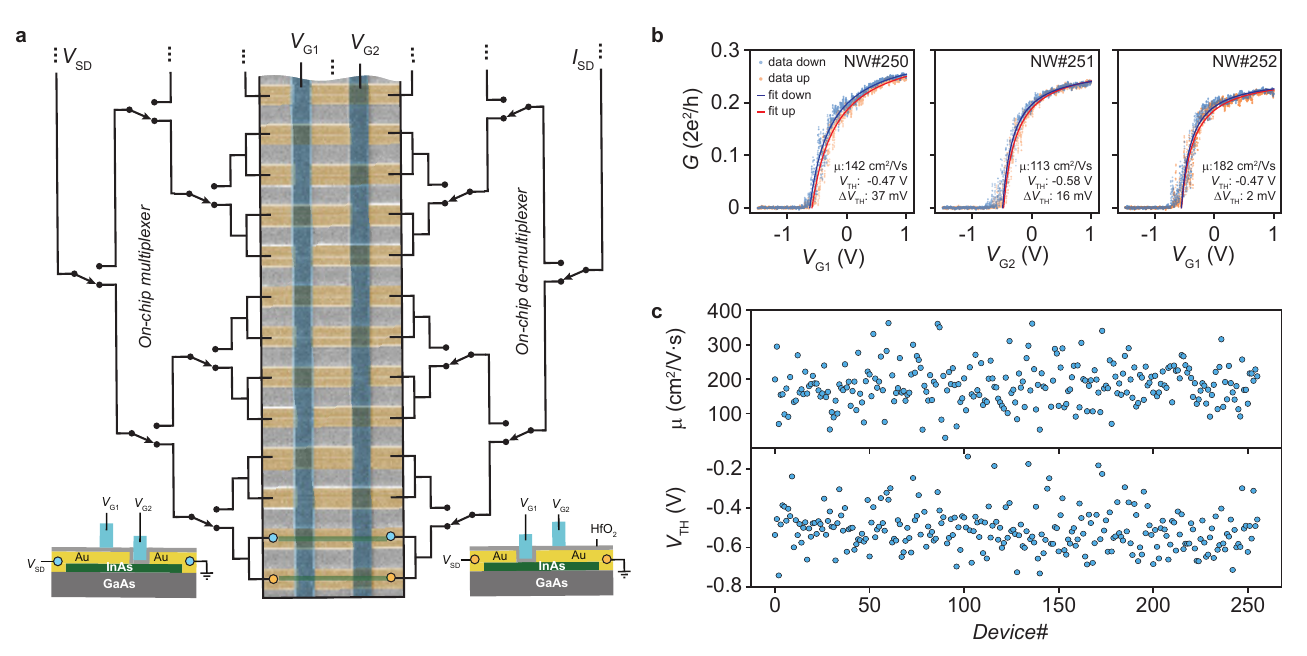} 
    \caption{\textbf{a} Schematic illustration of the multiplexer circuit used to address 256 individual SAG nanowire FET devices. The middle image shows a scanning electron micrograph of a part of the device array. Insets show schematic cross-sections of two neighboring devices. Note that the position of the exposed semiconductor segment alternates between devices, thus allowing the final selection to be carried out with the two top gates ($V_\mathrm{G1}$ and $V_\mathrm{G2}$). \textbf{b} Example of the conductance as a function of gate voltage for three consecutive devices in the array. Data for the gate sweep towards negative voltage (down) is shown in blue, and towards positive voltage (up) - in orange. Blue and red lines are fits to the expression $G^{-1} = R_s + L^2/(\mu_\mathrm{FE} C(V - V_\mathrm{TH})$. Fit parameters for the threshold voltage,$V_\mathrm{TH}$ and mobility, $\mu$, for the down traces are stated in each case, and panel \textbf{c} shows the extracted parameters for all 256 devices in the array. 
    } 
    \label{Fig3}
  \hfill
\end{figure*}

As an example, Fig.\ \ref{Fig2}\textbf{a} shows $H$ along the length of a single nanowire having an average height of $82.3 \, \mathrm{nm}$ and a standard deviation of $\pm 0.9 \, \mathrm{nm}$. 
The black curve shows the running average, $\overline{H}$, using a $1\,\mu \mathrm m$ averaging interval, relevant for comparison with $L_\mathrm{D} = 1 \, \mu \mathrm m$ FET devices as discussed below. The relative standard deviation of $\overline{H}$  along the length of the nanowire is $1.5\%$. To further quantify this and to investigate systematic variations also across the array, Fig.~\ref{Fig2}\textbf{b} shows $\overline H$ from the center of each of $180$ individual nanowires measured at positions along two directions across the array ($\Vec{v_1}$ and $\Vec{v_2}$ in Fig.\ref{Fig1}). $\overline H$ increases along both directions and linear fits yield slopes of $4.4$\,nm/mm and $4.2$\,nm/mm along $\Vec{v_1}$ and $\Vec{v_2}$, respectively. Assuming that the linearity extends across the entire array, a plane-fit of $\overline H(x,y)$ using these slopes yields an estimate of a maximum of $6\%$ variation of $\overline H$ across a distance of $1 \, \mathrm{mm}$ on the substrate. We attribute this variation to large-scale spatial modulation of the substrate temperature during growth\cite{madsen2011influence}. Such considerable variations could be important for integrated nanowire circuits covering a significant area. In addition to the overall trend observed in $\overline{H}$, variations between neighboring nanowires are clearly significant. To quantify this, Fig.\ \ref{Fig2}\textbf{c} shows histograms of $\overline{H}$ centered at $82.4 \pm 1.6$\,nm and the residuals of $\overline{H}$ after subtracting the linear trend along $\Vec{v_1}$ and $\Vec{v_2}$. The standard deviation $\sigma_\mathrm H = 0.97\,\mathrm{nm}~(1.1\%~\mathrm{of}~\overline{H})$ corresponds to the average small-range height variations between the nominally identical nanowires.


\begin{figure*}[htb]
  \centering
    \includegraphics[width = 16.5 cm]{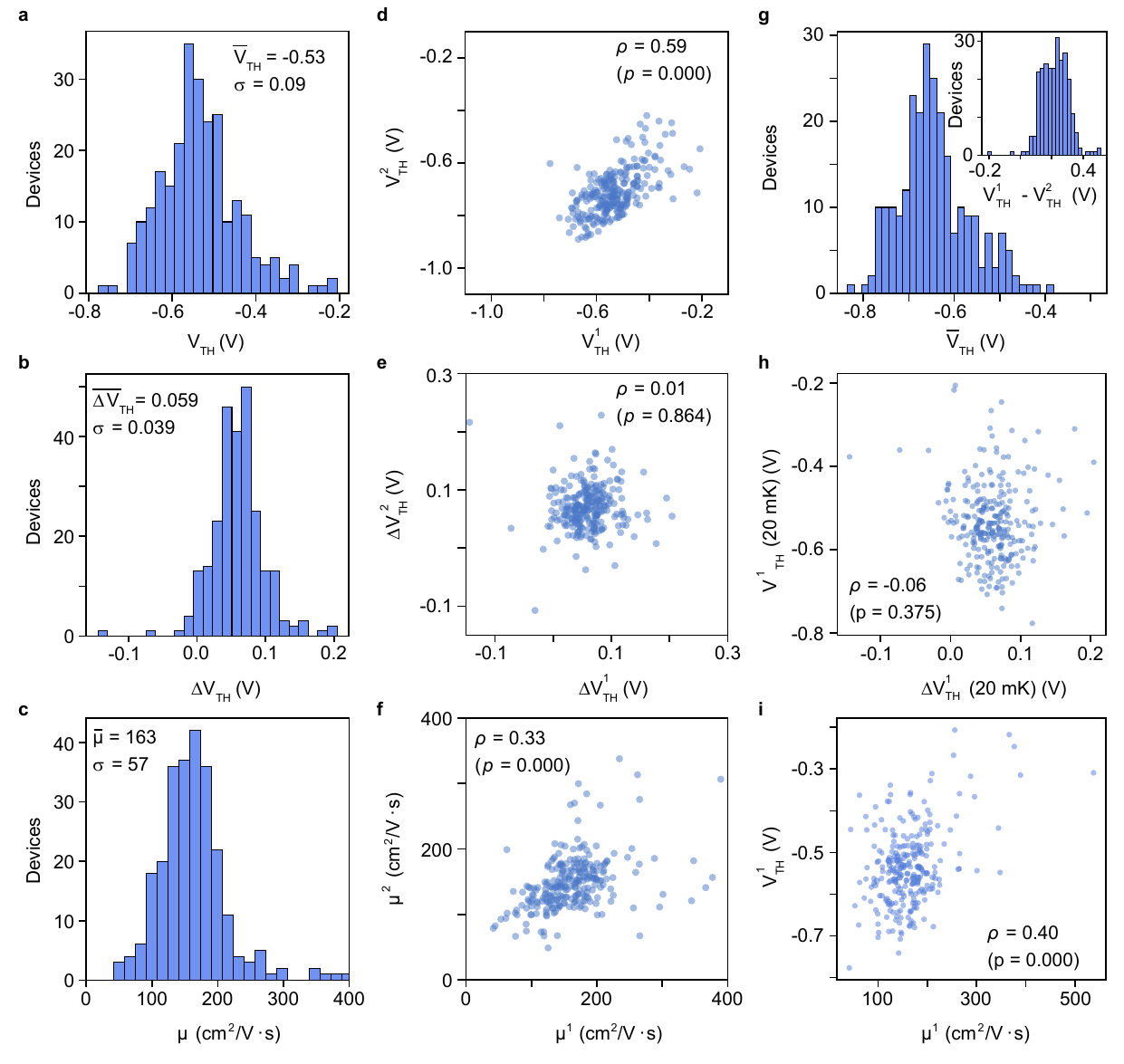}
    \caption{ \textbf{a,b,c} Histograms of threshold voltage $V_\mathrm{TH}$, hysteresis $\Delta V_\mathrm{TH}$, and mobility $\mu$, respectively, for 256 devices measured at $20~\mathrm{mK}$. \textbf{d,e,f} Correlations between two cool-downs to $20\ \mathrm{mK}$ for $V_\mathrm{TH}$, $\Delta{}V_\mathrm{TH}$, and $\mu{}_\mathrm{FE}$ respectively. Pearson's correlation coefficient $\rho$ and the corresponding \textit{p}-value shown in each panel. \textbf{g} Distribution of $\overline{V_\mathrm{TH}} = (V_\mathrm{TH}^1 + V_\mathrm{TH}^2)/2$ and the differences $(V_\mathrm{TH}^1 - V_\mathrm{TH}^2)$ of threshold values measured for the two cool-downs. \textbf{h} Correlation between $V_\mathrm{TH}$ and $\Delta V_\mathrm{TH}$ \textbf{i} Correlation between $V_\mathrm{TH}$ and $\mu$.
    }
    \label{Fig4}
  \hfill
\end{figure*}


Having discussed the structural variability across the array, we now consider the corresponding distribution and reproducibility of low-temperature electrical characteristics for nominally identical devices. A cryogenic, on-chip multiplexer/de-multiplexer circuit was used to enable the characterization of 256 individual, lithographically identical SAG nanowire FETs in a single cool-down (See Ref.\ \cite{olvsteins2023cryogenic} for details). Individual nanowires were electrically contacted by Ti/Au electrodes in an FET geometry keeping a constant channel length of $L_\mathrm{D}=1\,\mu \mathrm m$ and top-gates separated from the InAs channel by $15 \, \text{nm}$ of $\mathrm{HfO}_2$. Figure~\ref{Fig3}\textbf{a} shows an SEM micrograph of a part of the studied device array which is situated in the row indicated by the red arrow in Fig.~\ref{Fig1}\textbf{b}. The MUX circuit, which is shown schematically (an optical microscope image of the entire circuit is shown in Supplementary Section VIII.), addresses devices in pairs, and two gates ($V_\mathrm{G1}, V_\mathrm{G2}$) are used for the final selection; if one gate is active for a particular device, the other gate is inactive due to screening by the contact metal (insets to Fig.~\ref{Fig3}a). In the following, we denote the potential of the active gate of any device by $V_\mathrm{G}$.

Figure \ref{Fig3}\textbf{b} shows three representative examples of the conductance as a function of gate voltage, $G(V_\mathrm{G})$, measured at $T= 20 \, \mathrm{mK}$. Typical $n$-type depletion mode FET behavior is observed as expected for InAs nanowires. By fitting $G(V_\mathrm G)$ to the standard expression\cite{gul2015towards} $G^{-1} = R_s + L_\mathrm{D}^2/(\mu_\mathrm{FE} C(V - V_\mathrm{TH}))$ -- shown as orange lines in Fig.~\ref{Fig3}\textbf{b} -- we extract the field effect mobility, $\mu_\mathrm{FE}$, threshold voltage, $V_\mathrm{TH}$, and series resistance, $R_\mathrm{s}$. The series resistance $R_\mathrm{s}$ accounts for contact resistance and the resistance of the measurement circuit. Numerical simulation was used to estimate the gate capacitance $C = 5.3 \mathrm{fF}$\cite{olvsteins2023cryogenic}, which was used for all fits. Also stated in the figure are values for the difference in threshold voltage between positive $(up)$ and negative $(dwn)$ $V_\mathrm G$ sweep directions $\Delta{}V_\mathrm{TH} = V_\mathrm{TH}^\mathrm{dwn} - V_\mathrm{TH}^\mathrm{up}$, used here as a measure of hysteresis.

Since effects of quantum confinement and surface scattering decrease with increasing nanowire dimensions, lower $V_\mathrm{TH}$ and larger $\mu$ are expected~\cite{choi2001threshold, yuan2009analytic, granzner2007analytical,ford:2009, hou2013diameter, feng2017low} for larger $\overline H$. In Ref.\ \cite{Nagda2023SAG} we investigated this relationship for InAs SAG nanowire and found approximately linear relationships with $\partial V_\mathrm{TH}/\partial H = -14.1 \pm 1.0 \mathrm{mV/nm}$, and $\partial \mu/\partial H = 12.0 \pm 2.3 \mathrm{cm}^2/\mathrm{Vs \cdot nm}$. This connects the structural properties and geometric variability discussed above with the variability of the electrical characteristics. If the geometric variability in Fig.~\ref{Fig2} is dominating the statistics of the electrical properties, the standard deviations of the $\overline H$ distribution, $\sigma_\mathrm H = 0.97 \, \mathrm{nm}$ would lead to a spread in $V_\mathrm{TH}$ of $13.7\,\mathrm{mV}$ and in $\mu$ of  $11.6 \, \mathrm{cm^2/Vs}$.

Figure \ref{Fig3}\textbf{c} shows $\mu$ and $V_\mathrm{TH}$ extracted for all 256 devices as a function of the position in the array (see Supplementary Sections VI and VII for the underlying data and fitting procedure). Device \#1 and \#256 are separated on the chip by $\sim 0.5~\mathrm{mm}$ and the absence of any clear trend shows that either large scale, systematic structural variations (cf.\ Fig.~\ref{Fig2}\textbf{b}) are not significant in this array or that other effects dominate the electrical properties. Figure 4\textbf{a} shows the distribution of $V_\mathrm{TH}$, which provides important information about the gate ranges required for optimal operation large-scale SAG nanowire circuits. In the present case, all devices in the array are in pinch-off (accumulation) for $V_\mathrm{G} < -0.8\, \mathrm V$, ($V_\mathrm{G} > -0.2\, \mathrm V$), respectively. The FETs of the MUX circuit (cf.\ Fig.~\ref{Fig3} and Ref.\ \cite{olvsteins2023cryogenic}) were operated at $V_\mathrm G \pm 1 \, \mathrm V$ to ensure robust closed (open) states\cite{olvsteins2023cryogenic}.

The threshold voltage has a mean value of $\overline{V_\mathrm{TH}} \approx -530 \, \mathrm{mV}$ and a standard deviation $\sigma_\mathrm{V_{TH}} = 90 \, \mathrm{mV}$ ($\sim 17 \%$ of $\overline{V_\mathrm{TH}}$). The larger relative standard deviation of $V_\mathrm{TH}$ compared to those of $\overline H$ suggests significant contributions to the variability from sources not directly detectable through $H$. One example could be microscopic crystal defects, as discussed above, or the effects of scattering and screening from random, charged impurities in the vicinity of the device. The latter depends on material quality, the quality of oxides, surface adsorbents, the parameters of processing, and experimental conditions. The charged impurity configuration remains stable if the impurity energy depth exceeds the thermal energy and the energy scale of external electrical fields, and we, therefore, expect a static background at the millikelvin temperatures and $V_\mathrm G$ ranges used here. However, thermally cycling the device $20\,\mathrm{mK} \rightarrow 300\, \mathrm{K} \rightarrow 20\,\mathrm{mK}$ causes a random re-configuration of the charged impurities,\cite{see:2012,holloway:2013} allowing us to study their contribution to device reproducibility. 

Figure~\ref{Fig4}\textbf{d} shows the correlation of the two $V_\mathrm{TH}$ values measured for each device at base temperature before and after thermal cycling having a correlation coefficient of $\rho = 0.59$. This shows that - although not detectable through $H$ - fixed structural/intrinsic properties partly govern the transport parameters. At the same time, however, a significant scatter in Fig.~\ref{Fig4}\textbf{d} shows that charged impurities - randomized by thermal cycling - also contribute to the overall spread in the $V_\mathrm{TH}$. The corresponding distribution in Fig.~\ref{Fig4}\textbf{g} of the mean for each device estimates the structural distribution with a standard deviation $\sigma_{\mathrm{\overline{V_{TH}}}} = 103\,\mathrm{mV}$ which again exceeds the expectation from the variations of $\overline H$. We speculate that intrinsic defects such as atomic-scale crystal defects, stacking faults, and material intermixing could be responsible for these thermal cycling independent variations between devices, which do not show up in the distribution of the geometric parameter $\overline H$.

The mobility is influenced by the same structural properties, and the distribution (Fig.~\ref{Fig4}\textbf{c}) and correlations (Fig.~\ref{Fig4}\textbf{f}) show qualitatively similar behavior, albeit with a larger dispersion caused by impurity charges. The hysteresis, $\Delta V_\mathrm{TH}$, on the other hand, is linked to gate-activated charge traps near the nanowire channel and is thus expected to be less dependent on the structure. This is consistent with the absence of correlations between the values obtained before and after thermal cycling, as seen in Fig.~\ref{Fig4}\textbf{e}.

Fig.~\ref{Fig4}\textbf{h} shows $V_\mathrm{TH}$ vs.\ $\Delta V_\mathrm{TH}$ for all devices in the same cool-down. No correlations are observed, which is consistent with the hypothesis that the hysteresis is caused by rearranging surface charges - a process that is independent of the threshold voltage of the device. Similarly, Fig.~\ref{Fig4}\textbf{i} shows $V_\mathrm{TH}$ vs.\ $\mu$. Here, both properties are dependent on the structure, and correlation is observed, which indicates that the intrinsic features governing the variation of threshold voltage are related to those determining the mobility.

Combined, the results show that the reproducibility in the values of $\mu$ and $V_\mathrm{TH}$ is influenced approximately equally by fixed structural/intrinsic properties and by the distribution of charged impurities, which vary between cool-downs. Thus, to improve the reproducibility of SAG nanowire transport properties at low temperatures, these results suggest that efforts should target both contributions. The intrinsic contribution cannot be accounted for by variation in the physical dimensions of the SAG nanowires, and thus, variations in composition or crystal defects may be the main source.


In summary, we have presented a comprehensive study of the reproducibility of structural and electrical parameters of nominally identical InAs nanowires realized by selective area growth. Large arrays of nanowires were grown, and statistics on the structural properties were acquired by high-resolution AFM analysis of 180 nanowires systematically spaced across the array. We quantify the nanowire-to-nanowire variability caused by local fluctuations in growth dynamics, lithographic accuracy, and large-scale systematic trends, presumably due to spatial temperature fluctuations across the growth wafer. To acquire statistically significant distributions of electron mobilities, threshold voltages, and hysteresis, we employed an on-chip cryogenic multiplexer design, allowing independent characterization of 256 individual devices. Comparing the spread of values to that expected from the distribution of structural parameters and by correlating values between successive cool-downs to millikelvin temperatures, we quantify the contributions to the variability from intrinsic fixed properties and from charged impurities in the vicinity of the devices.

The $\sigma_\mathrm{V_{TH}}$ in our system is comparable to values found in AlGaAs/GaAs 2DEGs\cite{al2013cryogenic, yang2009potential} and foundry cryo-CMOS samples\cite{zwerver2022qubits, bavdaz2022quantum}. However, it must be noted that the width of the gates in our system is larger than those of Ref.\ \cite{zwerver2022qubits, bavdaz2022quantum}, and variability is expected to decrease with increasing gate size\cite{zwerver2022qubits} as the influence of impurities on the electrostatic environment averages over a larger scale\cite{kuhn2011process}. Nevertheless, given that this is the first study of reproducibility of InAs SAG and no specific attempts were made to optimize reproducibility in the studied structures, the results clearly show the potential of the platform. In further experiments intrinsic defects can be reduced by including an intermediary buffer layer\cite{beznasyuk:2022} and the amount of charged impurities can be reduced by employing in-situ or resist-free processing\cite{heedt:2021}.

The statistical approach developed here will be an important tool for optimizing SAG nanowires and other bottom-up nanomaterials towards large-scale circuits and may motivate efforts towards developing high throughput TEM methodologies enabling statistical correlations between electrical quantum properties and the atomic scale structures. Finally, we emphasize that concerning the SAG nanowires, even with the current, low-mobility structures, reproducibility is sufficient for realizing large-scale functional circuits such as multiplexer/de-multiplexer\cite{olvsteins2023cryogenic} and other logic circuitry. For circuits relying on high $\mu_\text{FE}$ or bandwidth, further optimization of $\mu_\text{FE}$ and the reproducibility of $\mu_\text{FE}$ is needed. The insights developed here will inform future optimization toward quantum devices with even stricter tolerances.

\subsection*{Acknowledgements}
This research was supported by research grants from Villum Fonden (Grant no.: 00013157) (TSJ) and the European Research Council under the European Union’s Horizon 2020 research and innovation programme (Grant no.: 716655 and 866158) (TSJ) and Microsoft Quantum. ICN2 acknowledges funding from Generalitat de Catalunya (Grant no.: 2021SGR00457) (JA). This study is part of the Advanced Materials programme and was supported by MCIN with funding from European Union NextGenerationEU (Grant no.: PRTR-C17.I1) (JA) and by Generalitat de Catalunya.  The authors thank support from “ERDF A way of making Europe”, by the “European Union”. ICN2 is supported by the Severo Ochoa program from Spanish MCIN / AEI (Grant no.: CEX2021-001214-S) and is funded by the CERCA Programme / Generalitat de Catalunya. Authors acknowledge the use of instrumentation as well as the technical advice provided by the National Facility ELECMI ICTS, node "Laboratorio de Microscopías Avanzadas" at University of Zaragoza. We acknowledge support from CSIC Interdisciplinary Thematic Platform (PTI+) on Quantum Technologies (PTI-QTEP+).

\subsection*{Competing interests}
The authors declare no competing interests.

\subsection*{Supplementary Information} Supplementary information contains extended information on structural characterisation and extended transport datasets.

\bibliographystyle{naturemag}
\bibliography{ref,TSJZotero}

\begin{thebibliography}{10}
\expandafter\ifx\csname url\endcsname\relax
  \def\url#1{\texttt{#1}}\fi
\expandafter\ifx\csname urlprefix\endcsname\relax\def\urlprefix{URL }\fi
\providecommand{\bibinfo}[2]{#2}
\providecommand{\eprint}[2][]{\url{#2}}

\bibitem{riel:2014}
\bibinfo{author}{Riel, H.}, \bibinfo{author}{Wernersson, L.-E.}, \bibinfo{author}{Hong, M.} \& \bibinfo{author}{{del Alamo}, J.~A.}
\newblock \bibinfo{title}{{{III}}\textendash{{V}} compound semiconductor transistors\textemdash from planar to nanowire structures}.
\newblock \emph{\bibinfo{journal}{MRS Bulletin}} \textbf{\bibinfo{volume}{39}}, \bibinfo{pages}{668--677} (\bibinfo{year}{2014}).

\bibitem{delalamo:2011}
\bibinfo{author}{Alamo, D.} \& \bibinfo{author}{A, J.}
\newblock \bibinfo{title}{Nanometre-scale electronics with {{III}}\textendash{{V}} compound semiconductors}.
\newblock \emph{\bibinfo{journal}{Nature}} \textbf{\bibinfo{volume}{479}}, \bibinfo{pages}{317--323} (\bibinfo{year}{2011}).

\bibitem{gonzalez-zalba:2021}
\bibinfo{author}{{Gonzalez-Zalba}, M.~F.} \emph{et~al.}
\newblock \bibinfo{title}{Scaling silicon-based quantum computing using {{CMOS}} technology}.
\newblock \emph{\bibinfo{journal}{Nat Electron}} \textbf{\bibinfo{volume}{4}}, \bibinfo{pages}{872--884} (\bibinfo{year}{2021}).

\bibitem{dick:2010}
\bibinfo{author}{Dick, K.~A.}, \bibinfo{author}{Thelander, C.}, \bibinfo{author}{Samuelson, L.} \& \bibinfo{author}{Caroff, P.}
\newblock \bibinfo{title}{Crystal {{Phase Engineering}} in {{Single InAs Nanowires}}}.
\newblock \emph{\bibinfo{journal}{Nano Lett.}} \textbf{\bibinfo{volume}{10}}, \bibinfo{pages}{3494--3499} (\bibinfo{year}{2010}).

\bibitem{ford:2009}
\bibinfo{author}{Ford, A.~C.} \emph{et~al.}
\newblock \bibinfo{title}{Diameter-{{Dependent Electron Mobility}} of {{InAs Nanowires}}}.
\newblock \emph{\bibinfo{journal}{Nano Lett.}} \textbf{\bibinfo{volume}{9}}, \bibinfo{pages}{360--365} (\bibinfo{year}{2009}).

\bibitem{bjork:2002}
\bibinfo{author}{Bj{\"o}rk, M.~T.} \emph{et~al.}
\newblock \bibinfo{title}{One-dimensional {{Steeplechase}} for {{Electrons Realized}}}.
\newblock \emph{\bibinfo{journal}{Nano Lett.}} \textbf{\bibinfo{volume}{2}}, \bibinfo{pages}{87--89} (\bibinfo{year}{2002}).

\bibitem{krogstrup:2015}
\bibinfo{author}{Krogstrup, P.} \emph{et~al.}
\newblock \bibinfo{title}{Epitaxy of semiconductor\textendash superconductor nanowires}.
\newblock \emph{\bibinfo{journal}{Nature Mater}} \textbf{\bibinfo{volume}{14}}, \bibinfo{pages}{400--406} (\bibinfo{year}{2015}).

\bibitem{Wang2019Jun}
\bibinfo{author}{Wang, N.} \emph{et~al.}
\newblock \bibinfo{title}{{Shape Engineering of InP Nanostructures by Selective Area Epitaxy}}.
\newblock \emph{\bibinfo{journal}{ACS Nano}} \textbf{\bibinfo{volume}{13}}, \bibinfo{pages}{7261--7269} (\bibinfo{year}{2019}).

\bibitem{op2020plane}
\bibinfo{author}{Op~het Veld, R.~L.} \emph{et~al.}
\newblock \bibinfo{title}{In-plane selective area {InSb}--{Al} nanowire quantum networks}.
\newblock \emph{\bibinfo{journal}{Communications Physics}} \textbf{\bibinfo{volume}{3}}, \bibinfo{pages}{1--7} (\bibinfo{year}{2020}).

\bibitem{Raya2020Jan}
\bibinfo{author}{Raya, A.~M.} \emph{et~al.}
\newblock \bibinfo{title}{{GaAs nanoscale membranes: prospects for seamless integration of III{\textendash}Vs on silicon}}.
\newblock \emph{\bibinfo{journal}{Nanoscale}} \textbf{\bibinfo{volume}{12}}, \bibinfo{pages}{815--824} (\bibinfo{year}{2020}).

\bibitem{Bollani2020Jan}
\bibinfo{author}{Bollani, M.} \emph{et~al.}
\newblock \bibinfo{title}{{Selective Area Epitaxy of GaAs/Ge/Si Nanomembranes: A Morphological Study}}.
\newblock \emph{\bibinfo{journal}{Crystals}} \textbf{\bibinfo{volume}{10}}, \bibinfo{pages}{57} (\bibinfo{year}{2020}).

\bibitem{beznasyuk:2022}
\bibinfo{author}{Beznasyuk, D.~V.} \emph{et~al.}
\newblock \bibinfo{title}{Doubling the mobility of {{InAs}}/{{InGaAs}} selective area grown nanowires}.
\newblock \emph{\bibinfo{journal}{Phys. Rev. Materials}} \textbf{\bibinfo{volume}{6}}, \bibinfo{pages}{034602} (\bibinfo{year}{2022}).

\bibitem{Friedl2018Apr}
\bibinfo{author}{Friedl, M.} \emph{et~al.}
\newblock \bibinfo{title}{{Template-Assisted Scalable Nanowire Networks}}.
\newblock \emph{\bibinfo{journal}{Nano Lett.}} \textbf{\bibinfo{volume}{18}}, \bibinfo{pages}{2666--2671} (\bibinfo{year}{2018}).

\bibitem{Lee2019Aug}
\bibinfo{author}{Lee, J.~S.} \emph{et~al.}
\newblock \bibinfo{title}{{Selective-area chemical beam epitaxy of in-plane InAs one-dimensional channels grown on InP(001), InP(111)B, and InP(011) surfaces}}.
\newblock \emph{\bibinfo{journal}{Phys. Rev. Mater.}} \textbf{\bibinfo{volume}{3}}, \bibinfo{pages}{084606} (\bibinfo{year}{2019}).

\bibitem{seidl2021postgrowth}
\bibinfo{author}{Seidl, J.} \emph{et~al.}
\newblock \bibinfo{title}{Postgrowth shaping and transport anisotropy in two-dimensional inas nanofins}.
\newblock \emph{\bibinfo{journal}{ACS nano}} \textbf{\bibinfo{volume}{15}}, \bibinfo{pages}{7226--7236} (\bibinfo{year}{2021}).

\bibitem{vaitiekenas2018selective}
\bibinfo{author}{Vaitiek{\.e}nas, S.} \emph{et~al.}
\newblock \bibinfo{title}{Selective-area-grown semiconductor-superconductor hybrids: A basis for topological networks}.
\newblock \emph{\bibinfo{journal}{Physical Review Letters}} \textbf{\bibinfo{volume}{121}}, \bibinfo{pages}{147701} (\bibinfo{year}{2018}).

\bibitem{Krizek2018Sep}
\bibinfo{author}{Krizek, F.} \emph{et~al.}
\newblock \bibinfo{title}{{Field effect enhancement in buffered quantum nanowire networks}}.
\newblock \emph{\bibinfo{journal}{Phys. Rev. Mater.}} \textbf{\bibinfo{volume}{2}}, \bibinfo{pages}{093401} (\bibinfo{year}{2018}).

\bibitem{hertel:2022}
\bibinfo{author}{Hertel, A.} \emph{et~al.}
\newblock \bibinfo{title}{Gate-{{Tunable Transmon Using Selective-Area-Grown Superconductor-Semiconductor Hybrid Structures}} on {{Silicon}}}.
\newblock \emph{\bibinfo{journal}{Phys. Rev. Appl.}} \textbf{\bibinfo{volume}{18}}, \bibinfo{pages}{034042} (\bibinfo{year}{2022}).

\bibitem{goswami:2023}
\bibinfo{author}{Goswami, A.} \emph{et~al.}
\newblock \bibinfo{title}{Sn/{{InAs Josephson Junctions}} on {{Selective Area Grown Nanowires}} with in {{Situ Shadowed Superconductor Evaporation}}}.
\newblock \emph{\bibinfo{journal}{Nano Lett.}} \textbf{\bibinfo{volume}{23}}, \bibinfo{pages}{7311--7318} (\bibinfo{year}{2023}).

\bibitem{ten2022large}
\bibinfo{author}{Ten~Kate, S.~C.} \emph{et~al.}
\newblock \bibinfo{title}{Small charging energies and g-factor anisotropy in pbte quantum dots}.
\newblock \emph{\bibinfo{journal}{Nano Letters}} \textbf{\bibinfo{volume}{22}}, \bibinfo{pages}{7049--7056} (\bibinfo{year}{2022}).

\bibitem{smith:2014}
\bibinfo{author}{Smith, L.~W.} \emph{et~al.}
\newblock \bibinfo{title}{Statistical study of conductance properties in one-dimensional quantum wires focusing on the 0.7 anomaly}.
\newblock \emph{\bibinfo{journal}{Phys. Rev. B}} \textbf{\bibinfo{volume}{90}}, \bibinfo{pages}{045426} (\bibinfo{year}{2014}).

\bibitem{smith2020high}
\bibinfo{author}{Smith, L.~W.} \emph{et~al.}
\newblock \bibinfo{title}{High-throughput electrical characterization of nanomaterials from room to cryogenic temperatures}.
\newblock \emph{\bibinfo{journal}{ACS nano}} \textbf{\bibinfo{volume}{14}}, \bibinfo{pages}{15293--15305} (\bibinfo{year}{2020}).

\bibitem{olvsteins2023cryogenic}
\bibinfo{author}{Ol{\v{s}}teins, D.} \emph{et~al.}
\newblock \bibinfo{title}{Cryogenic multiplexing using selective area grown nanowires}.
\newblock \emph{\bibinfo{journal}{Nature Communications}} \textbf{\bibinfo{volume}{14}}, \bibinfo{pages}{7738} (\bibinfo{year}{2023}).

\bibitem{see:2012}
\bibinfo{author}{See, A.~M.} \emph{et~al.}
\newblock \bibinfo{title}{Impact of {{Small-Angle Scattering}} on {{Ballistic Transport}} in {{Quantum Dots}}}.
\newblock \emph{\bibinfo{journal}{Phys. Rev. Lett.}} \textbf{\bibinfo{volume}{108}}, \bibinfo{pages}{196807} (\bibinfo{year}{2012}).

\bibitem{schroer:2010}
\bibinfo{author}{Schroer, M.~D.} \& \bibinfo{author}{Petta, J.~R.}
\newblock \bibinfo{title}{Correlating the {{Nanostructure}} and {{Electronic Properties}} of {{InAs Nanowires}}}.
\newblock \emph{\bibinfo{journal}{Nano Lett.}} \textbf{\bibinfo{volume}{10}}, \bibinfo{pages}{1618--1622} (\bibinfo{year}{2010}).

\bibitem{madsen2011influence}
\bibinfo{author}{Madsen, M.~H.}, \bibinfo{author}{Aagesen, M.}, \bibinfo{author}{Krogstrup, P.}, \bibinfo{author}{S{\o}rensen, C.} \& \bibinfo{author}{Nyg{\aa}rd, J.}
\newblock \bibinfo{title}{Influence of the oxide layer for growth of self-assisted inas nanowires on si (111)}.
\newblock \emph{\bibinfo{journal}{Nanoscale research letters}} \textbf{\bibinfo{volume}{6}}, \bibinfo{pages}{1--5} (\bibinfo{year}{2011}).

\bibitem{gul2015towards}
\bibinfo{author}{G{\"u}l, {\"O}.} \emph{et~al.}
\newblock \bibinfo{title}{Towards high mobility insb nanowire devices}.
\newblock \emph{\bibinfo{journal}{Nanotechnology}} \textbf{\bibinfo{volume}{26}}, \bibinfo{pages}{215202} (\bibinfo{year}{2015}).

\bibitem{choi2001threshold}
\bibinfo{author}{Choi, Y.-K.}, \bibinfo{author}{Ha, D.}, \bibinfo{author}{King, T.-J.} \& \bibinfo{author}{Hu, C.}
\newblock \bibinfo{title}{Threshold voltage shift by quantum confinement in ultra-thin body device}.
\newblock In \emph{\bibinfo{booktitle}{Device Research Conference. Conference Digest (Cat. No. 01TH8561)}}, \bibinfo{pages}{85--86} (\bibinfo{organization}{IEEE}, \bibinfo{year}{2001}).

\bibitem{yuan2009analytic}
\bibinfo{author}{Yuan, Y.}, \bibinfo{author}{Yu, B.}, \bibinfo{author}{Song, J.} \& \bibinfo{author}{Taur, Y.}
\newblock \bibinfo{title}{An analytic model for threshold voltage shift due to quantum confinement in surrounding gate mosfets with anisotropic effective mass}.
\newblock \emph{\bibinfo{journal}{Solid-State Electronics}} \textbf{\bibinfo{volume}{53}}, \bibinfo{pages}{140--144} (\bibinfo{year}{2009}).

\bibitem{granzner2007analytical}
\bibinfo{author}{Granzner, R.}, \bibinfo{author}{Schwierz, F.} \& \bibinfo{author}{Polyakov, V.~M.}
\newblock \bibinfo{title}{An analytical model for the threshold voltage shift caused by two-dimensional quantum confinement in undoped multiple-gate mosfets}.
\newblock \emph{\bibinfo{journal}{IEEE Transactions on Electron Devices}} \textbf{\bibinfo{volume}{54}}, \bibinfo{pages}{2562--2565} (\bibinfo{year}{2007}).

\bibitem{hou2013diameter}
\bibinfo{author}{Hou, J.~J.} \emph{et~al.}
\newblock \bibinfo{title}{Diameter dependence of electron mobility in {InGaAs} nanowires}.
\newblock \emph{\bibinfo{journal}{Applied Physics Letters}} \textbf{\bibinfo{volume}{102}}, \bibinfo{pages}{093112} (\bibinfo{year}{2013}).

\bibitem{feng2017low}
\bibinfo{author}{Feng, W.}, \bibinfo{author}{Peng, C.}, \bibinfo{author}{Li, S.} \& \bibinfo{author}{Li, X.-Q.}
\newblock \bibinfo{title}{Low-field electron mobility of {InSb} nanowires: Numerical efforts to larger cross sections}.
\newblock \emph{\bibinfo{journal}{Scientific Reports}} \textbf{\bibinfo{volume}{7}}, \bibinfo{pages}{1--8} (\bibinfo{year}{2017}).

\bibitem{Nagda2023SAG}
\bibinfo{author}{Nagda, G.} \emph{et~al.}
\newblock \bibinfo{title}{Dimension dependent electrical transport in selective area grown inas nanowires}.
\newblock \emph{\bibinfo{journal}{To be submitted}}  (\bibinfo{year}{2024}).

\bibitem{holloway:2013}
\bibinfo{author}{Holloway, G.~W.}, \bibinfo{author}{Song, Y.}, \bibinfo{author}{Haapamaki, C.~M.}, \bibinfo{author}{LaPierre, R.~R.} \& \bibinfo{author}{Baugh, J.}
\newblock \bibinfo{title}{Trapped charge dynamics in {{InAs}} nanowires}.
\newblock \emph{\bibinfo{journal}{Journal of Applied Physics}} \textbf{\bibinfo{volume}{113}}, \bibinfo{pages}{024511} (\bibinfo{year}{2013}).

\bibitem{al2013cryogenic}
\bibinfo{author}{Al-Taie, H.} \emph{et~al.}
\newblock \bibinfo{title}{Cryogenic on-chip multiplexer for the study of quantum transport in 256 split-gate devices}.
\newblock \emph{\bibinfo{journal}{Applied Physics Letters}} \textbf{\bibinfo{volume}{102}}, \bibinfo{pages}{243102} (\bibinfo{year}{2013}).

\bibitem{yang2009potential}
\bibinfo{author}{Yang, Q.-Z.}, \bibinfo{author}{Kelly, M.}, \bibinfo{author}{Farrer, I.}, \bibinfo{author}{Beere, H.} \& \bibinfo{author}{Jones, G.}
\newblock \bibinfo{title}{The potential of split-gate transistors as one-dimensional electron waveguides revealed through the testing and analysis of yield and reproducibility}.
\newblock \emph{\bibinfo{journal}{Applied Physics Letters}} \textbf{\bibinfo{volume}{94}}, \bibinfo{pages}{033502} (\bibinfo{year}{2009}).

\bibitem{zwerver2022qubits}
\bibinfo{author}{Zwerver, A.} \emph{et~al.}
\newblock \bibinfo{title}{Qubits made by advanced semiconductor manufacturing}.
\newblock \emph{\bibinfo{journal}{Nature Electronics}} \textbf{\bibinfo{volume}{5}}, \bibinfo{pages}{184--190} (\bibinfo{year}{2022}).

\bibitem{bavdaz2022quantum}
\bibinfo{author}{Bavdaz, P.} \emph{et~al.}
\newblock \bibinfo{title}{A quantum dot crossbar with sublinear scaling of interconnects at cryogenic temperature}.
\newblock \emph{\bibinfo{journal}{npj Quantum Information}} \textbf{\bibinfo{volume}{8}}, \bibinfo{pages}{86} (\bibinfo{year}{2022}).

\bibitem{kuhn2011process}
\bibinfo{author}{Kuhn, K.~J.} \emph{et~al.}
\newblock \bibinfo{title}{Process technology variation}.
\newblock \emph{\bibinfo{journal}{IEEE Transactions on Electron Devices}} \textbf{\bibinfo{volume}{58}}, \bibinfo{pages}{2197--2208} (\bibinfo{year}{2011}).

\bibitem{heedt:2021}
\bibinfo{author}{Heedt, S.} \emph{et~al.}
\newblock \bibinfo{title}{Shadow-wall lithography of ballistic superconductor\textendash semiconductor quantum devices}.
\newblock \emph{\bibinfo{journal}{Nat Commun}} \textbf{\bibinfo{volume}{12}}, \bibinfo{pages}{4914} (\bibinfo{year}{2021}).

\end{thebibliography}


\begin{thebibliography}{1}
\expandafter\ifx\csname url\endcsname\relax
  \def\url#1{\texttt{#1}}\fi
\expandafter\ifx\csname urlprefix\endcsname\relax\def\urlprefix{URL }\fi
\providecommand{\bibinfo}[2]{#2}
\providecommand{\eprint}[2][]{\url{#2}}

\bibitem{olvsteins2023cryogenic}
\bibinfo{author}{Ol{\v{s}}teins, D.} \emph{et~al.}
\newblock \bibinfo{title}{Cryogenic multiplexing using selective area grown nanowires}.
\newblock \emph{\bibinfo{journal}{Nature Communications}} \textbf{\bibinfo{volume}{14}}, \bibinfo{pages}{7738} (\bibinfo{year}{2023}).

\bibitem{Isu1998Jun}
\bibinfo{author}{Isu, T.}, \bibinfo{author}{Hata, M.}, \bibinfo{author}{Watanabe, A.} \& \bibinfo{author}{Katayama, Y.}
\newblock \bibinfo{title}{{In situ scanning microprobe reflection high{-}energy electron diffraction observation of GaAs surfaces during molecular{-}beam epitaxial growth}}.
\newblock \emph{\bibinfo{journal}{J. Vac. Sci. Technol., B}} \textbf{\bibinfo{volume}{7}}, \bibinfo{pages}{714} (\bibinfo{year}{1998}).

\bibitem{Beznasyuk2022Mar}
\bibinfo{author}{Beznasyuk, D.~V.} \emph{et~al.}
\newblock \bibinfo{title}{{Doubling the mobility of InAs/InGaAs selective area grown nanowires}}.
\newblock \emph{\bibinfo{journal}{Phys. Rev. Mater.}} \textbf{\bibinfo{volume}{6}}, \bibinfo{pages}{034602} (\bibinfo{year}{2022}).

\bibitem{LER}
\bibinfo{author}{Dinh, C.~Q.}
\newblock \bibinfo{title}{Lacerm}.
\newblock \bibinfo{howpublished}{https://www.lacerm.com/home}.

\bibitem{gul:2015}
\bibinfo{author}{G{\"u}l, {\"O}.} \emph{et~al.}
\newblock \bibinfo{title}{Towards high mobility {{InSb}} nanowire devices}.
\newblock \emph{\bibinfo{journal}{Nanotechnology}} \textbf{\bibinfo{volume}{26}}, \bibinfo{pages}{215202} (\bibinfo{year}{2015}).

\end{thebibliography}

\subsection*{Data Availability}
The data supporting the findings is available at https://doi.org/10.11583/DTU.24967755

\newpage
\begin{figure*}[htb]
  \centering
    \includegraphics[width = \linewidth]{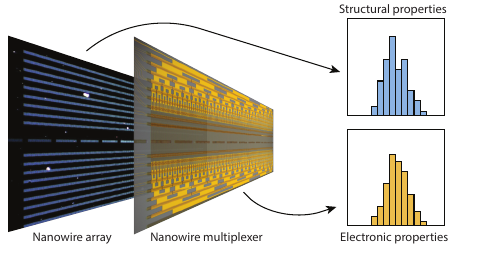}
    
    \label{TOC Graphic}
  \hfill
\end{figure*}

\end{document}


\author{Dags Olsteins*}
\affiliation{Department of Energy Conversion and Storage, Technical University of Denmark, 2800 Kgs.Lyngby, Denmark}

\author{Gunjan Nagda*}
\affiliation{Center For Quantum Devices, Niels Bohr Institute, University of Copenhagen,\\ 2100 Copenhagen, Denmark}

\author{Damon J. Carrad}
\affiliation{Department of Energy Conversion and Storage, Technical University of Denmark, 2800 Kgs.Lyngby, Denmark}

\author{Daria V. Beznasyuk}
\affiliation{Department of Energy Conversion and Storage, Technical University of Denmark, 2800 Kgs.Lyngby, Denmark}

\author{Christian E. N. Petersen}
\affiliation{Department of Energy Conversion and Storage, Technical University of Denmark, 2800 Kgs.Lyngby, Denmark}

\author{Sara Martí-Sánchez}
\affiliation{Catalan Institute of Nanoscience and Nanotechnology (ICN2), CSIC and BIST, Campus UAB, 08193 Bellaterra, Barcelona, Catalonia, Spain}

\author{Jordi Arbiol}
\affiliation{Catalan Institute of Nanoscience and Nanotechnology (ICN2), CSIC and BIST, Campus UAB, 08193 Bellaterra, Barcelona, Catalonia, Spain}
\affiliation{ICREA, Passeig de Lluís Companys 23, 08010 Barcelona, Catalonia, Spain}

\author{Thomas Sand Jespersen}
\affiliation{Department of Energy Conversion and Storage, Technical University of Denmark, 2800 Kgs.Lyngby, Denmark}
\affiliation{Center For Quantum Devices, Niels Bohr Institute, University of Copenhagen,\\ 2100 Copenhagen, Denmark}
\email{tsaje@dtu.dk}

\title{Supplementary Information \\ Statistical Reproducibility of Selective Area Grown InAs Nanowire Devices}

\maketitle 

\section{Growth of selective area grown (SAG) nanowires}

The experiments performed in this study are based on the nanowires from the same sample as used in Ref.~\cite{olvsteins2023cryogenic}. 

After preparing the substrate for selective area growth, the sample is introduced to the MBE chamber. The sample is then degassed in the loadlock for 4 hours at $200$°C and then transferred to the tunnel connecting the loadlock to the main chamber. Here, the sample is further degassed at a heating station for 1 hour at 400\,°C. Prior to MBE growth, the substrate is thermally annealed by increasing the substrate temperature to $T_{sub} = 400$°C with a ramp rate of $20$°C/min and then further to $T_{sub} = 620$°C with a ramp rate of $10$°C/min under a constant As$_4$ beam equivalent over-pressure of $1.4 \times 10^{-5}$\,mbar. We use reflection high energy electron diffraction (RHEED) to track the removal of the native oxide from the exposed substrate at this temperature for approximately $13$ minutes (this time varies slightly from sample to sample), the last minute of which the intensity of the specular spot does not increase\cite{Isu1998Jun}. Subsequently, $T_{sub}$ is reduced to the GaAs(Sb) buffer growth temperature of $T_{{sub},{GaAs(Sb)}} = 600$°C. For the buffer growth, a As/Ga ratio of $9$ and Sb/Ga ratio of $3$ is used. This ensures that the growth rate is dependent on the Ga flux and the growth is performed with a Ga growth rate of $0.1$\,ML/s for $30$ min. With these conditions Sb acts as a surfactant to assist in the growth of well-defined facets and Sb is not incorporated into the final structures. The substrate temperature is further reduced to $T_{{sub},{InAs}} = 500$°C for InAs growth. The substrate temperature is monitored before, shortly at the beginning and after the growth using a pyrometer and the variation in $T_{sub}$ is maximally $2$°C. This uncertainty is attributed to the variation in temperature read-out caused by radiation from different regions of the substrate. An As/In ratio of $9$ and an In growth rate of $0.06$\,ML/s is used. The choice of lower a $T_{sub}$ and a slower growth rate is employed to minimize temperature assisted Ga diffusion into the InAs channel\cite{Beznasyuk2022Mar}. 

\newpage
\section{Mask Opening Etch Roughness}

\begin{figure}[htb]
  \centering
    \includegraphics[width = \linewidth]{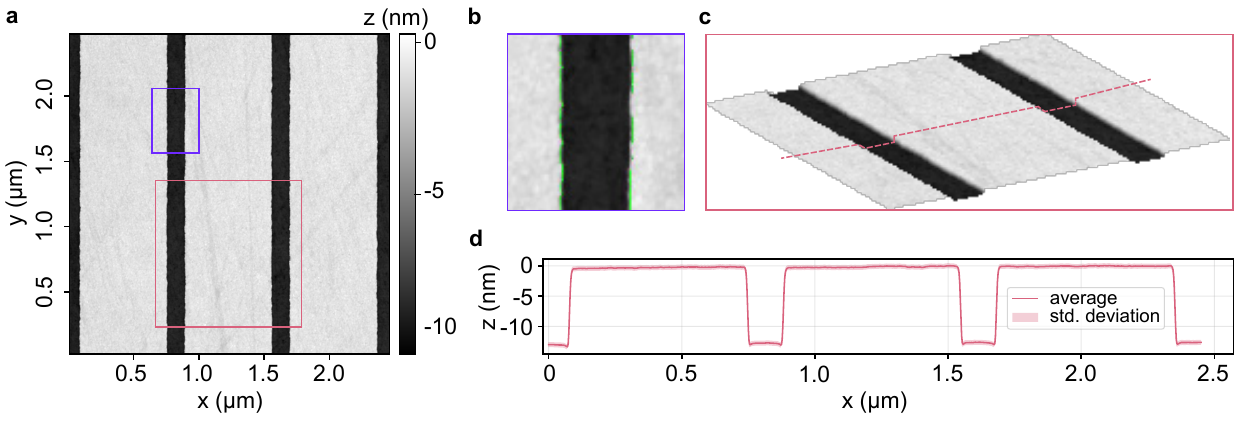}
    \caption{\textbf{a} AFM micrograph of a substrate prepared for SAG. \textbf{b} Zoom-in on the area in the purple box in \textbf{a}. Green lines show the edges of the trench. \textbf{c} A 3D render of the area in the red box in \textbf{a}. \textbf{d} Averaged linecuts across \textbf{a} and the standard deviation.
    } 
    \label{Supp_prefab}
  \hfill
\end{figure}

Figure~\ref{Supp_prefab}(a) shows a $2.5 \times 2.5\, \mu \text{m}$ AFM micrograph of a substrate prepared for SAG, representative of the sample used for nanowire growth. AFM is typically performed on the substrates prior to SAG to ensure cleanliness, determine the line edge roughness (LER) of the trenches along with the trench width. The LER is calculated using an open source software - Lacerm\cite{LER} - which detects the edges of the trenches (see green lines in Fig.~\ref{Supp_prefab}(b)) with the image contrast, and in this case is determined to be 1.08 nm. A close-up 3D view of the AFM micrograph is presented in Fig.~\ref{Supp_prefab}(c) showing a clean $\mathrm{SiO}_2$ mask surface as well as the trenches etched in the mask. Figure~\ref{Supp_prefab}(d) depicts the line profiles averaged over $2.5\, \mu \text{m}$ (solid line) and the corresponding standard deviation as a coloured band. 

\newpage
\section{Crystal structure and Strain}

\begin{figure}[htb]
  \centering
    \includegraphics[width = \linewidth]{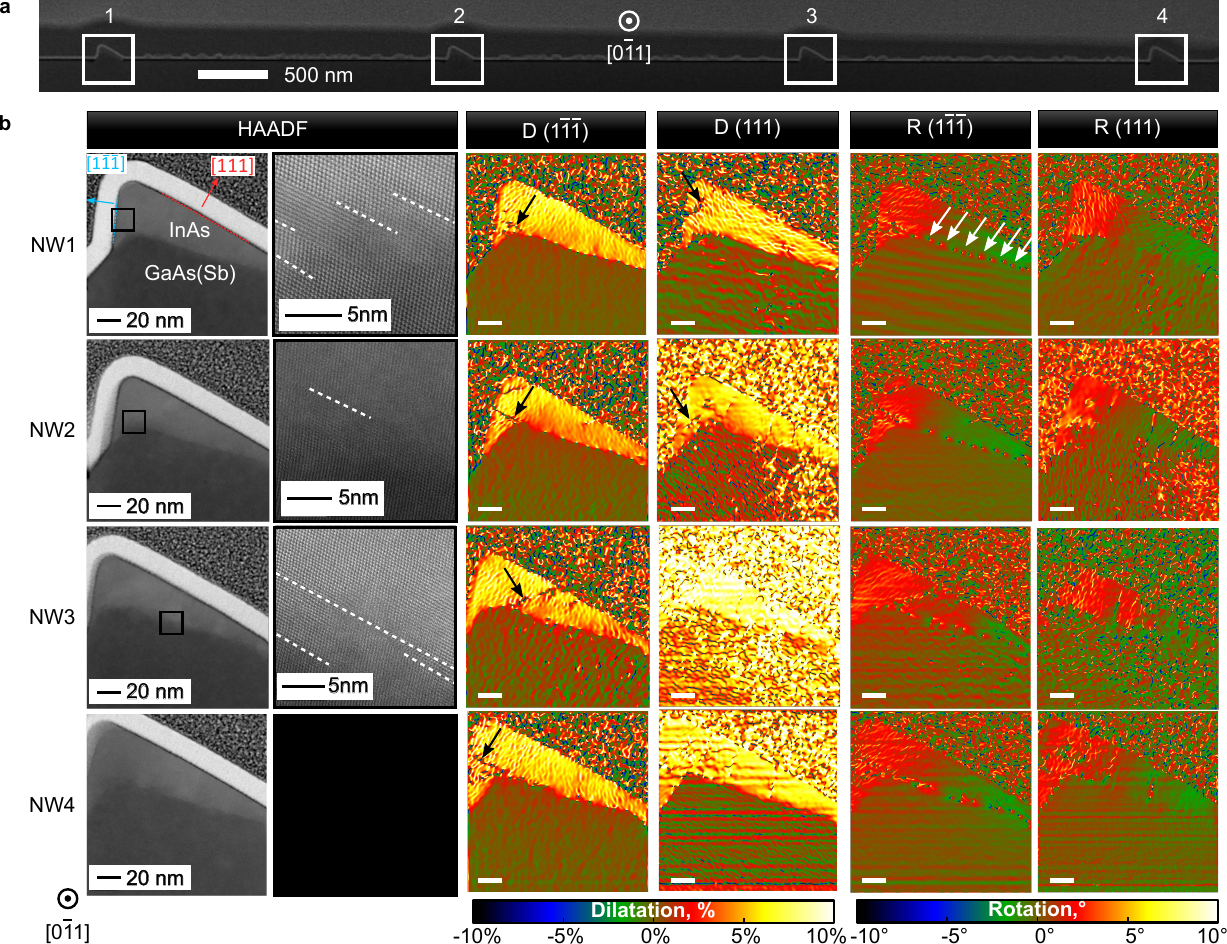}
    \caption{\textbf{a} A cross-sectional view of 4 InAs/GaAs(Sb) nanowires roughly from the middle of an array of 50 nanowires oriented along the \hkl[0-11] direction on a GaAs\hkl(311)A substrate. \textbf{b} (column 1) High-angle annular dark field scanning tunneling micrographs (HAADF-STEM) show the GaAs(Sb) and InAs regions of the nanowires. The morphology of all 4 nanowires is uniform and the InAs channel exhibits \hkl{111}A facets. \textbf{b} (column 2) Stacking faults originating at the GaAs(Sb)/InAs interface are highlighted with white dashed lines. \textbf{b} (column 3, 4) Dilatation maps of all nanowires show stacking faults, indicated by black arrows. \textbf{b} (column 5, 6) The rotational maps indicate rotation of the crystal planes between the GaAs(Sb) buffer and the InAs channel, an array of dislocations at the GaAs(Sb)/InAs interface are indicated with white arrows for nanowire1. Scale bars correspond to $20$\,nm in all dilatation and rotation maps.
    } 
    \label{Supp_HAADF}
  \hfill
\end{figure}

Figure~\ref{Supp_HAADF} shows the morphology, strain and crystal structure of the nanowires comparable to nanowires investigated in this study. A uniform morphology is visible for 4 InAs/GaAs(Sb) nanowires oriented along the \hkl[0-11] direction on a GaAs\hkl(311)A substrate in Fig~\ref{Supp_HAADF}a. The InAs channel exhibits \hkl{111}A facets. The cross-sectional shape is specific to the choice of substrate, in-plane orientation, mask dimensions such as the width and pitch, as well as the growth parameters listed in the previous section. 

Figure~\ref{Supp_HAADF}b shows the crystal structure of all 4 nanowires, along with the corresponding geometric phase analysis (GPA) of the dilatation and rotation maps. Figure~\ref{Supp_HAADF}b (column 2) shows stacking faults (highlighted as white dashed lines, and black arrows) originating at the GaAs(Sb)/InAs interface and propagating towards the InAs channel. It can be observed the dilatation maps in Fig.~\ref{Supp_HAADF}b (column 3, 4) that relaxation also occurs through creation of an array of dislocations at the GaAs(Sb)/InAs interface. To understand if the strain is fully relaxed via formation of misfit dislocations (indicated by white arrows), rotational maps are also plotted in Fig.~\ref{Supp_HAADF}b (column 5, 6). The \hkl{111}A planes show a splitting visible as a color change, indicating the presence of an asymmetric in-plane rotation at the nanowire edges. The rotation is stronger ($\sim3$°) on the shorter \hkl{111}A planes covering the more inclined regions of the GaAs(Sb) buffer than on the other side that sits atop the longer facet. Rotation on the side covering the large GaAs(Sb) facet is $\sim 0.65$°. The relaxation mechanism involves the formation of an array of misfit dislocations (plastic relaxation) and asymmetric elastic plane rotation the nanowires.

\newpage
\section{Elemental composition}

\begin{figure}[htb]
  \centering
    \includegraphics[width = 0.4 \linewidth]{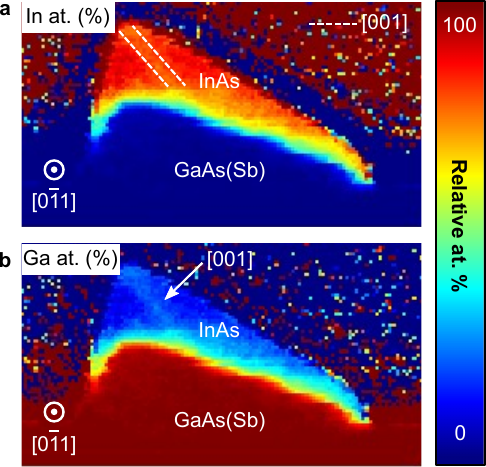}
    \caption{\textbf{a} In and \textbf{b} Ga relative atomic \% compositional map of a typical nanowire used in this study obtained by electron energy loss spectroscopy (EELS). The InAs region is found to be diluted with $8-10\%$ Ga. The white dashed lines in \textbf{a} show the Ga-rich region within the InAs channel along which Ga is found to diffuse more than in the surrounding region. The preferential diffusion of Ga along the \hkl[001] direction is better visible in \textbf{b} where the map is relative to the Ga atomic \%.
    } 
    \label{Supp_HAADF-2}
  \hfill
\end{figure}

Electron Energy Loss Spectroscopy (EELS) of the nanowires was used to obtain the elemental composition of the nanowire. In particular, relative elemental quantification of the In vs Ga ratio in atomic \% is shown in  Fig.~\ref{Supp_HAADF-2}a and b respectively. Ga diffusion towards the InAs channel of ~8-10 \% was observed along the \hkl[001] direction. The diffusion region is highlighted in Fig.~\ref{Supp_HAADF-2}a between the white dashed lines and is better visible in Fig.~\ref{Supp_HAADF-2}b. Such preferential diffusion has been reported in GaAs(Sb)/InGaAs/InAs nanowires grown on GaAs\hkl(001) substrates \cite{Beznasyuk2022Mar}. It has been proposed that Ga diffusion arises as a thermally activated strain minimization mechanism during growth of lattice-mismatched InAs and GaAs. An In rich region only exists at the outermost layers of the conduction channel whereas the region in between is diluted with the diffused Ga. This can be circumvented by further reducing the substrate temperature during InAs growth, but may compromise the selectivity of the sample.

\clearpage 
\section{Details of AFM Data Analysis}

\begin{figure}[htb]
  \centering
    \includegraphics[width=\linewidth]{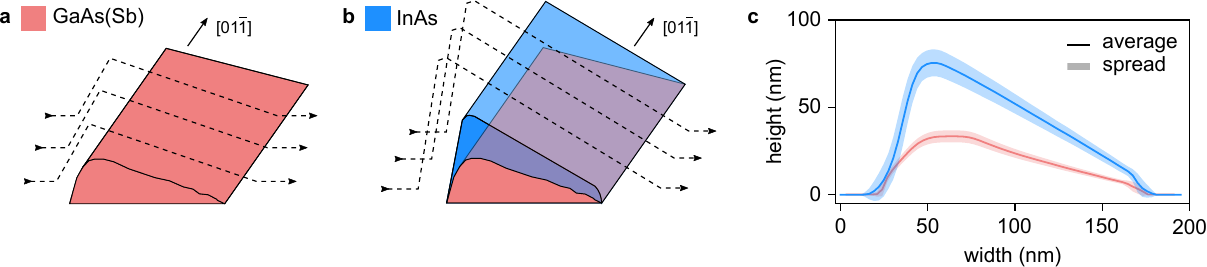}
    \caption{\textbf{a} Schematic of a \hkl[01-1]-oriented GaAs(Sb) nanowire grown on a reference sample used to measure the cross-sectional profile of the GaAs(Sb) buffer layer. The dotted line traces indicate the AFM scan lines which were used along a length of 1$\mu$um of the nanowire to obtain the average profile. \textbf{b} Schematic of a \hkl[01-1]-oriented nanowire representative of the InAs nanowires used for the electrical characterization. The dotted line traces indicate the AFM scan lines which were used along a length of 1$\mu$um of the nanowire to obtain the average profile. \textbf{c} AFM profiles of the GaAs(Sb) nanowires from the reference sample as well as the InAs nanowires, where the solid lines represent the average profile of the nanowires and the shaded region corresponds to the standard deviation calculated from AFM data along a length of 1$\mu$m.
    } 
    \label{AFM-analysis}
  \hfill
\end{figure}

The nanowires used for this study were grown on the same substrate, in the same geometry as the nanowires used for the electrical characterization and are part of the same growth. In order to check if there is a systematic dependence of the nanowire geometry on the location in the arrays, we measure a set of $5$ nanowires from each of the $18$ rows approximately along a straight (diagonal) line as indicated by the vectors in main text Fig.\ 1. Only 1$\mu$m long regions approximately in the center of the 10$\mu$m long nanowires are considered since this corresponds to the channel length of the nanowireFET devices used in the multiplexer circuits and the nanowireFET statistics. We also measure $2$ nanowires from a reference sample which consists of only the GaAs(Sb) buffer layer along a length of 1$\mu$m. The resolution of all AFM data shown corresponds to $100$ scan lines over a length of 1$\mu$m of the nanowires per image and $1024$ points per scan line. The dashed lines in Fig.~\ref{AFM-analysis}(a) and (b) are representative of the AFM scan lines on GaAs(Sb) and InAs nanowires. Using all the nanowire profiles (scan lines), we calculate the average profile as well as the standard deviation along the 1$\mu$m long segments of all nanowires as seen in Fig.~\ref{AFM-analysis}(a). The GaAs(Sb) nanowires show a more consistent profile as can be seen in the small standard deviation. This is not unexpected since the homoepitaxial growth of GaAs(Sb) on a GaAs substrate should result in a more homogeneous growth when compared to the growth of lattice-mismatched InAs, which shows a larger spread. A Python script was developed to extract the height ($H$) of all measured nanowires.\\

Considering the large-scale variations of the nanowire height the linear fits for $H$ measured at positions along $\Vec{v_1}$ and $\Vec{v_2}$ yields slopes of $4.4$\,nm/mm and $4.2$\,nm/mm, respectively (cf.\ Fig.~1b and Fig.~2 of the main manuscript). With these trends projected along the two directions, and assuming a linear dependence across the entire array we get $H(x,y) = (2.67\cdot 10^{-6})x - (4.17 \cdot 10^{-6})y + 7.99 \cdot 10^{-8}$ where $x,y$ are the nanowire positions using the co-ordinate system in Fig.~1\textbf{b} of the main manuscript. The magnitude of the gradient of this plane $|\nabla H| \approx 5 \cdot 10^{-6}$ corresponds to a $\sim 6\%$ variation in $H$ across a distance of $1 \, \text{mm}$ along the substrate.



\clearpage 
\section{Full Dataset of nanowireFET Device Measurements}

\begin{figure}[htb]
  \centering
    \includegraphics[width = 0.9\linewidth]{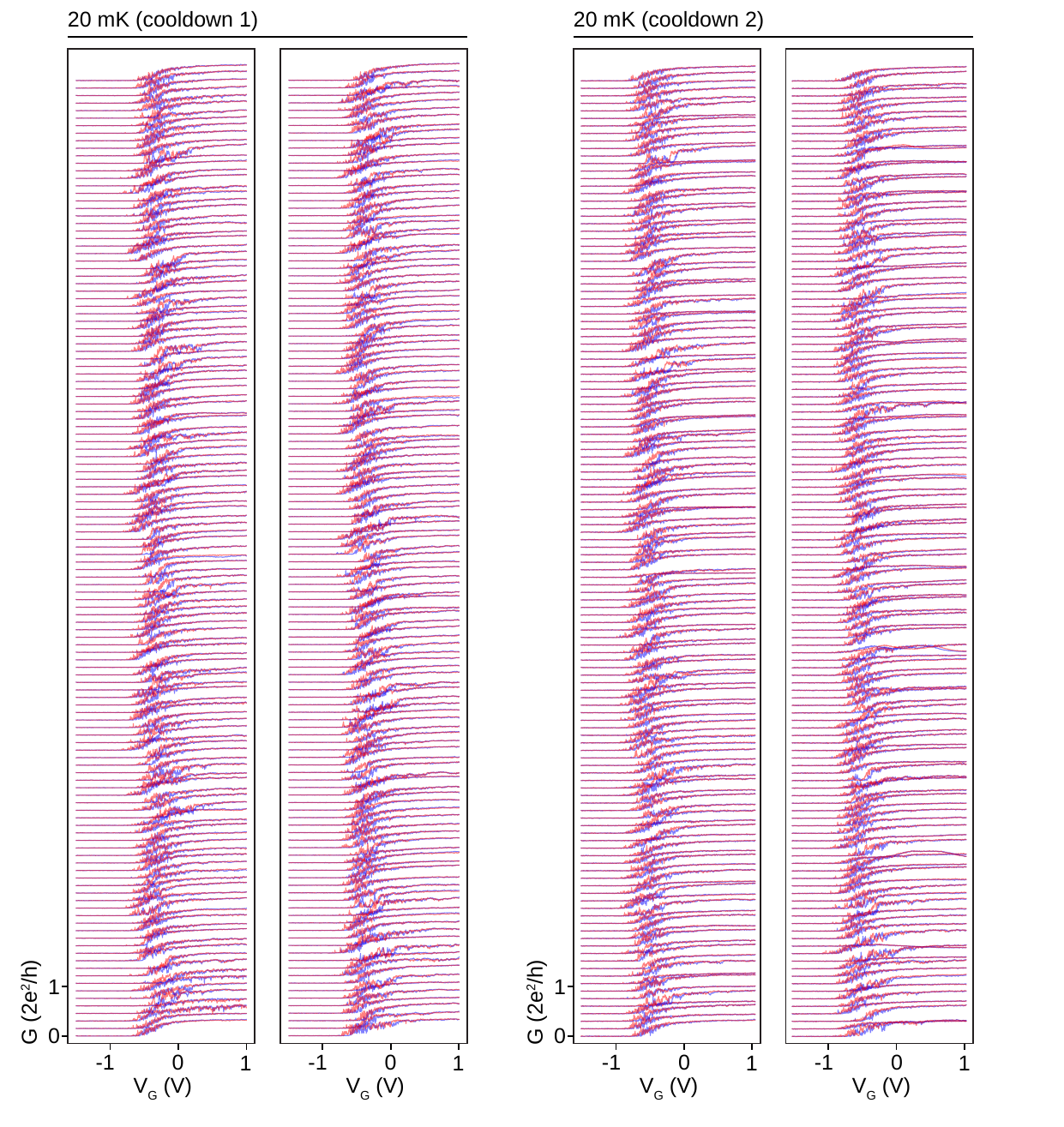}
    \caption{Extended data set of the measurements shown in main text Fig.\ 3 and 4. Raw data of pinch-off curves of 256 lithographically identical transistor devices, traces offset in conductance for clarity. Two rounds of measurements at 20 mK in two separate cooldowns. Gate voltage swept from negative (positive) to positive (negative) in red (blue).
    } 
    \label{Supp_stats}
  \hfill
\end{figure}

Figure \ref{Supp_stats} Shows the full data set of the measurements shown in main text Fig.4 consisting of pinch-off curves for 256 lithographically identical nanowireFET devices. The devices used are part of the multiplexer structure discussed in details in Ref.\ \cite{olvsteins2023cryogenic}.

\clearpage 
\section{Fitting Procedures}

To deal with the large number of devices, all transfer curves were analyzed using the same semi-automatic procedure. The fit range for each measurement is chosen by first smoothing the pinch-off curve using Savitzky-Golay filtering and then identifying the gate voltage at which the conductance rises to 5\% of its maximum value, the non-filtered data in the range from the identified gate voltage to the largest applied positive gate voltage is then used for fitting to the expression\cite{gul:2015} $G^{-1} = R_s + L^2/(\mu_\mathrm{FE} C(V - V_\mathrm{TH}))$. Estimating the gate capacitance which directly influences the absolute value of the extracted $\mu{}_\mathrm{FE}$ is challenging. To estimate the capacitance between the nanowire and the gate electrode, ANSYS electrostatic modelling software was used. The device geometry was reconstructed in the model using measurements from device SEM micrographs as well as AFM and cross-section TEM imaging of nanowires from the same growth. The gate dielectric (HfO$_2$) was estimated to be 15 nm thick and assumed to cover the device evenly and have a relative dielectric permittivity $\epsilon{}$ of 18.
The fit then yields the values for $V_\mathrm{TH}$, $\mu{}_\mathrm{FE}$, and $R_\mathrm{s}$. The values are used for relative comparisons between devices and any systematic biases of the model are assumed to be the same across devices and hence do not affect the relative relationships of extracted values between individual devices.

\clearpage 
\section{Mobility and threshold voltage for both cooldowns}

\begin{figure}[htb]
  \centering
    \includegraphics[width = 0.7\linewidth]{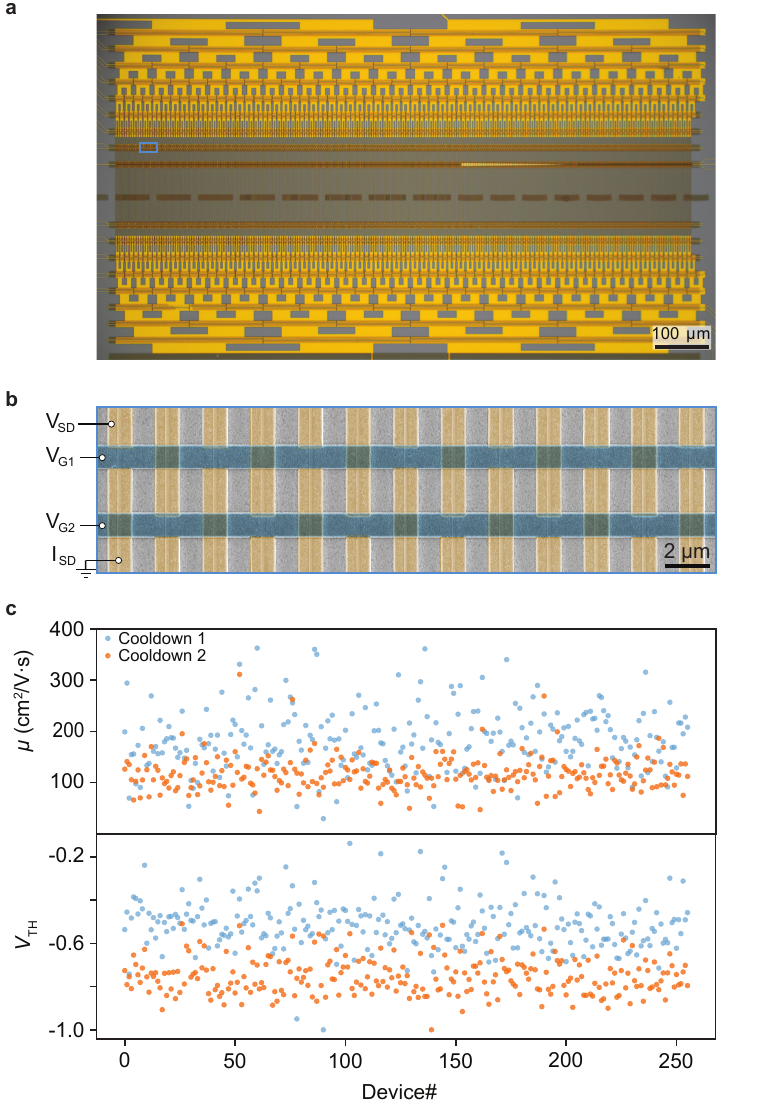}
    \caption{\textbf{a} Optical microscope micrograph of the multiplexer circuit used for transport measurements. \textbf{b} False-colored SEM micrograph of the nanowire FET devices in the blue rectangle in \textbf{a}. Contacts in gold, gates in blue, only one gate is active for each device. \textbf{c} $\mu$ and $V_\mathrm{TH}$ as function of device number for both cooldowns.
    } 
    \label{Supp_outs}
  \hfill
\end{figure}

The entire on-chip measurement circuit and the data behind the correlations in the main text Fig.~4. Figure~\ref{Supp_outs}a shows the SAG nanowire multiplexer, and Fig.~\ref{Supp_outs}c shows $\mu_\text{FE}$ and $V_\text{TH}$ as a function of device number for both cooldowns.

\section*{References}

\bibliographystyle{naturemag}
\bibliography{ref,TSJZotero}